\documentclass[osajnl,twocolumn,showpacs,superscriptaddress,10pt]{revtex4-1} 
\usepackage{amsmath,amssymb,graphicx}
\usepackage{color}
\usepackage{soul}
\usepackage{psfrag}

\begin{document}
\title{Ground-state entanglement of spin-1 bosons undergoing superexchange interactions in optical superlattices}
\author{Artur Barasi\'nski}
\author{Wies{\l}aw Leo\'nski}
\affiliation{Quantum Optics and Engineering Division, Institute of Physics, University of Zielona G\'ora, Z. Szafrana 4a, 65-516 Zielona G\' ora, Poland}
\author{Tomasz Sowi\'nski}
\affiliation{Institute of Physics of the Polish Academy of Sciences, Al. Lotnik\'ow 32/46, 02-668 Warsaw, Poland}
\affiliation{Center for Theoretical Physics of the Polish Academy of Sciences, Al. Lotnik\'ow 32/46, 02-668 Warsaw, Poland}

\begin{abstract}
We discuss a model with ultra-cold atoms 
confined in optical superlattices. In particular, we study the ground-state properties of two spin-1 bosons trapped in a double-well potential. Depending on the external magnetic field 
and biquadratic interactions different phases of magnetic order are realized. Applying von Neumann entropy and number of relevant orbitals, 
we quantify the bipartite entanglement between particles.  Changing the values of the parameters determining superlattices, we can switch the system  between 
differently entangled states.
\end{abstract}

\maketitle

\section{Introduction}
Quantum entanglement is one of fundamental concepts in physics and it plays a crucial role in quantum information theory 
\cite{bennetnatere2000}. Without idea of entanglement, it is not possible to understand many spectacular  properties of quantum systems or phenomena related to them. On can mention at this point teleportation \cite{boschiprl1998,bouwmeesternaure1997,M05,OKB07,GK13}, quantum cryptography 
\cite{gisinrmp2002,BLC13}, topological order \cite{levineprl2004,kitaevprl2006} or quantumness of physical systems \cite{BMW11} \textit{etc}. 
In a context of strongly correlated many-body systems, together with quantum statistics, the entanglement between indistinguishable 
particles is responsible for nonintuitive properties of macroscopic systems \cite{ghoshnature2003}. The relation between the ground-state 
entanglement and the quantum phase  transitions provide a bridge between quantum information theory and condensed matter physics 
\cite{osterlohnature2002,osbornepra2002,amicormp2008}. For instance, transitions from the product states to strongly entangled ones in a 
two-\textit{qubit} Heisenberg nuclear spin chain have been experimentally observed in systems affected by varying external magnetic field 
\cite{pengpra2005,zhangprl2008}. 

In recent years, a wide perspective for experimental realization of the spin chains systems have been found in models involving ultra-cold 
atoms trapped in optical lattices \cite{leweoxford2012,blattnatphys2012}, ultracold quantum gates \cite{blochnatphys2012}, or coupled 
quantum dots \cite{losspra1998} \textit{etc}. Contrary to the solid-state physics, ultra-cold atomic systems can be quite well tuned to 
situations in which their properties are well described by simplified models of condensed matter theory. Moreover, they permit to engineer 
and manipulate quantum states in single lattice sites as well as collectively on a whole lattice. It is possible to tune all relevant 
parameters almost adiabatically, even through quantum phase transitions \cite{greinernature2002, blochrevmodphys2008} and in consequence, to 
test scaling invariance and universality \cite{hungnature2011}. Such systems have also opened completely new possibilities for answering to 
fundamental questions of quantum information theory related to transitions from the product to entangled states, quantum state transfers, 
quantum magnetism, spin dynamics, \textit{etc}. \cite{nikolopoulosepl65_2004, christandlprl92_2004, wupra80_2009, weitenbergnature2011, 
pietraszkiewiczpra88_2013}. In this way, ultra-cold atomic systems become dedicated quantum simulators \cite{Feynman,Hauke} for condensed 
matter problems, where fundamental parameters of theoretical models can be controlled experimentally.

Properties of interacting ultra-cold atoms confined in optical lattices are typically described in the language of Hubbard-like models. In 
the simplest case of spinless bosons interacting repulsively via short-range delta-like interactions, the Bose-Hubbard model has only two 
competing terms related to the single-particle tunneling to the neighboring site and to the additional energy cost when two bosons meet in 
given site \cite{jakschprl81_1998}. Extended Hubbard models (for fermions and bosons) originate in taking into account higher bands of 
optical lattices, long-range interactions, and internal structure of interacting particles 
\cite{PhysRevLett.95.030405,PhysRevLett.95.033003,PhysRevA.85.033638,PhysRevLett.111.215302,PhysRevLett.108.165301}. In particular, 
scenarios when tunneling is dominated by interactions, all Hubbard models can be perturbatively simplified and can be rewritten to 
appropriate lattice-spin models. For example, when the standard Bose-Hubbard model is considered and the system remains in the Mott-insulator 
phase, one can find the most relevant corrections by treating the tunneling as a virtual process in the second order of perturbation theory 
\cite{auerbach1994}. With this observation, Simon \textit{et al.} mimicked experimentally the one-dimensional Ising model 
\cite{simonsnature2011}. In more general case, when internal structure of ultra-cold atoms can not be neglected, the low energy effective 
Hamiltonian requires superexchange interactions \cite{andersonpr1959}. Then, the corresponding spin model also includes higher powers of the  
scalar products of spin operators at neighboring sites, $H=\sum_k J_k (\hat{\boldsymbol{S}}_i \cdot \hat{\boldsymbol{S}}_j)^k$ 
\cite{imambekovpra68_2003, yipplr2003,wumplb20_2006, eckertnjp9_2007, hermeleprl103_2009}. For example, for the case of spin-1 bosons the 
system is described by the Heisenberg model with additional biquadratic term, and is called Quadratic-Biquadratic Heisenberg (QBH) model 
\cite{garciaprl2004,yipplr2003,imambekovpra68_2003}. So far, mostly the linear Heisenberg spin-models were intensively studied over wide 
range of parameters \cite{DL94,DD95,zhoupra2003,hiranojpsj2007,pendpra2010,guoepjd2010}. The QBH model have attracted great interest mainly due to a 
wealth of possible quantum phases, predicted in its antiferromagnetic case, and also in \textit{i.e.} Haldane, nematic, dimerized, trimerized phases 
\textit{etc}. \cite{rodrigezprl2011,chiaraprb2011,yippra85_2012}, as well. The QBH model was also successfully applied in the studies concerning 
magnetic properties and energy levels distribution in several real materials \cite{millet1999, louprl2000, basterdisprb2007, benciniica2008, 
semenkainchem2010}, where the validity of the biquadratic interaction has been emphasized.

Motivated by all these observations, we shall discuss here the properties of spin-1 bosons (for example alkali atoms of 23-sodium or 87-rubidium) confined in optical superlattice, \textit{i.e.} the lattice created by interference of two independent laser standing waves with commensurate 
frequencies ratio equal to $2$. Such lattice is characterized by a structure involving weakly coupled double-well potentials and therefore, it could be a quite good arena for 
studying two-site Hubbard models. Moreover, in the limit of strong repulsions this model can be simplified to the two-site QBH model 
\cite{garciaprl2004,yipplr2003,imambekovpra68_2003}. We should note that in \cite{wagnerpra84_2011} the authors have concentrated on the effect of 
the asymmetry of the double-well potential in the realization of various types of magnetic order. They also discussed some aspects related 
to the bipartite entanglement between sites in a quite limited regime of parameters. Here, we complement and extend these studies assuming 
that full range of parameters of resulting Hamiltonian can be reached experimentally. It can be done directly (for instance by using the optical Feshbach 
resonances \cite{rodrigezprl2011, papoularpra81_2010}) or by performing some kind of ``quantum simulation'' where the system was prepared in its
excited state (like it was proposed \cite{garciaprl2004, chiaraprb2011}). In particular, our main goal is to analyze the influence of the 
biquadratic interaction to the ground-state entanglement. 

The paper is organized as follows. In Sec. II we introduce the model studied here and present its theoretical background. In Sec. III we give 
a full analysis of the spectrum of two-site Hamiltonian, and in Sec. IV we present phase diagram of the ground-state in a whole accessible 
range of parameters. Sec. V is devoted to the presentation of main results related to the entanglement generation in the ground state. Finally, we present our conclusions in Sec. V.

\section{Two-site spin-model Hamiltonian}
In this article, we consider spin-1 ultra-cold bosons confined in a one-dimensional superlattice potential. We assume that dynamics is frozen in two 
perpendicular directions and the particles remain in the ground state of confining potential. Moreover, that laser beams are configured in such a way that one can treat system as independent double-well potentials. In 
other words, we assume that tunneling amplitudes between potential dimmers are much smaller than the tunnelings inside the dimer. Such 
situation is well described with two-side Bose-Hubbard model of the form
\begin{align}
H &= -t (\hat{a}_{L\sigma}^\dagger \hat{a}_{R\sigma} + \hat{a}_{R\sigma}^\dagger\hat{a}_{L\sigma})+ 
\frac{U_0}{2} \sum_{i=L,R} \hat{n}_i(\hat{n}_i-1) + \nonumber \\ 
&+ \frac{U_2}{2} \sum_{i=L,R} \left(\hat{\boldsymbol{S}}_i^2 -2 \hat{n}_i\right) -~\gamma\boldsymbol{B}\cdot
\left(\hat{\boldsymbol{S}}_L+\hat{\boldsymbol{S}}_R\right), 
\label{HamHub}
\end{align}
where $\hat{a}_{i\sigma}$ annihilates boson in spin state $\sigma\in \{-1,0,1\}$ at left ($L$) or right ($R$) well, 
$\hat{n}_i = \sum_\sigma\hat{a}^\dagger_{i\sigma}\hat{a}_{i\sigma}$ is a total number of particles at give site, whereas  
$\hat{\boldsymbol{S}}_L$ and $\hat{\boldsymbol{S}}_R$ are total spin operators. The parameters $U_0$ and $U_2$ describe repulsive interaction between two particles confined in one lattice site.  They are proportional to the s-wave scattering lengths $a_{0}$ and $a_{2}$ when total spin of colliding particles is $S=0$ and $S=2$, respectively (see for example \cite{yipplr2003}). 
The last term in the Hamiltonian \eqref{HamHub} describes 
linear Zeeman effect in the external magnetic field $\boldsymbol{B}$. The parameter $\gamma=g\mu_B$, where $\mu_B$ is the Bohr magneton and 
$g$ is the Land\'e factor. In further analysis we will assume that external magnetic field is oriented along $z$ axis. It is worth to notice 
that inversion of the spin-axis quantization is equivalent to the inversion of the magnetic field. Therefore, without loosing generality, 
one can assume that magnetic field $B^z$ is not negative. All predictions for $B^z<0$ have their counterparts with opposite sign of the spins in 
$B^z>0$. In fact, the Hamiltonian  \eqref{HamHub} is an effective Hamiltonian, general enough to be applied in description of a large family of physical systems. For instance, it was used in discussion of various interesting phenomena appearing in Kerr-like quantum-optical or nano-
systems models, such as photon (phonon) blockade \cite{LMG10, GSK12, MPL13} (sometimes called as \textit{nonlinear quantum scissors} 
\cite{LT94,ML04,KL10,LK11}) or quantum-chaotic behavior \citep{M90,MH91,L96,KKL09,GSC13}.

Our aim is to study the system with unit filling in the strongly repulsive regime, \textit{i.e.} when non-local tunneling term in the Hamiltonian 
\eqref{HamHub} can be treated as a small perturbation (when compared with sum of local interaction terms). Therefore, we rewrite the 
Hamiltonian as the effective one in the second order of perturbation in $t$ \cite{yipplr2003,imambekovpra68_2003}:
\begin{multline} 
\label{eq:ham0}
 H = J_0 + J_1\hat{\boldsymbol{S}}_{L}\cdot\hat{\boldsymbol{S}}_{R} + J_{2}(\hat{\boldsymbol{S}}_{L}\cdot\hat{\boldsymbol{S}}_{R})^2 
-~\gamma B^z(\hat{S}^z_{L}+\hat{S}^z_{R}),
\end{multline}	
where $J_1=-\frac{2t^2}{U_2}$, $J_2=-\frac{2t^2}{3U_2}-\frac{4t^2}{3U_0}$ and $J_0=J_1-J_2$.

Since we shall  study effects of the biquadratic term in the Hamiltonian \eqref{eq:ham0}, it is convenient to parametrize the Hamiltonian 
\eqref{eq:ham0} with three dimensionless
parameters
\begin{equation} 
\label{eq:ham}
 H = \lambda\hat{\boldsymbol{S}}_{L}\cdot\hat{\boldsymbol{S}}_{R}+ \tan\theta(\hat{\boldsymbol{S}}_{L}\cdot\hat{\boldsymbol{S}}_{R})^2 
-~h(\hat{S}^z_{L}+\hat{S}^z_{R}),
\end{equation}
where $\lambda=J_1/|J_1|=\pm 1$ is a sign of a linear term, $\tan\theta=J_2/|J_1|$, and $h=\gamma  B^z/|J_1|$. For convenience, we also set 
the energy scale in such a way that $J_0=0$. Positive (negative) $\lambda$ favors antiferromagnetic (ferromagnetic) orientation of 
spins. It should be emphasized that although, from the model point of view, the angle $\theta$ lies within the interval $(-\pi/2,\pi/2)$, for the system of ultracold atoms when the ratio is typically $U_2/U_0>0$, angle $\theta$ can not cover this range completely
\cite{rodrigezprl2011}. Nevertheless, the experimental scheme where a whole range of $\theta$ can be obtained was also proposed 
\cite{garciaprl2004,chiaraprb2011}. Moreover, in contrast to usual condensed matter systems (see for example \cite{andersonpr1959}), ultra 
cold atoms make it possible to engineer experimental setups where biquadartic coupling is larger than linear one.

It should be also underlined that the effective Hamiltonian \eqref{eq:ham} can be derived from the more general Hubbard-like Hamiltonian 
\eqref{HamHub} only for repulsive interactions. Only for this case, the Mott-Insulator phase with one particle in each lattice site is a true 
ground-state of the system, in the limit of vanishing tunneling. This means that only negative sign of $\lambda$ can be obtained in this 
framework. However, it was shown recently that positive $\lambda$ can be effectively engineered by preparing system in its excited state 
\cite{garciaprl2004, chiaraprb2011}. In consequence, although such system is not in the true ground-state, it can be quite well described with use of the 
effective Hamiltonian \eqref{eq:ham} with positive $\lambda$ and for long time-scales. 

\section{Spectrum of the effective Hamiltonian}
\begin{figure}
\includegraphics{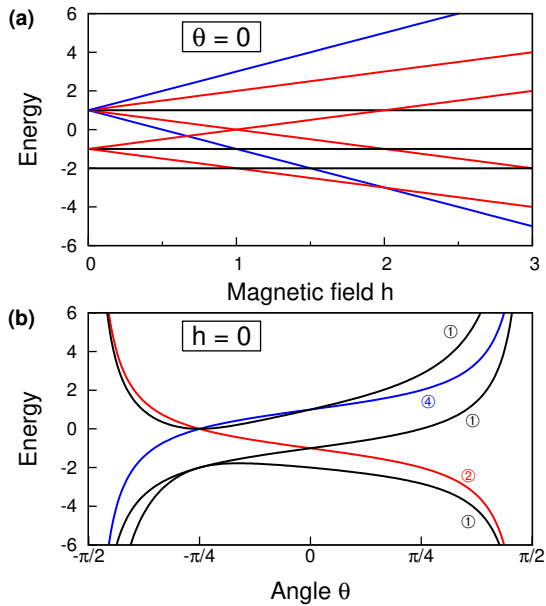}
\caption{Spectrum of the Hamiltonian \eqref{eq:ham} for $\lambda=1$ in two limiting situations: (a) vanishing biquadratic interaction 
$\theta=0$, (b) vanishing external magnetic field $h=0$.  In (a) different colors corresponds to states with different total spin $S^z$ ($S^z=0$ 
black line, $S^z=\pm 1$ red line, $S^z=\pm 2$ blue line); in (b) circled numbers denote degree of the degeneracy of the states.}
\label{fig:fig1}
\end{figure}

The Hamiltonian \eqref{eq:ham} commutes with the total spin operator $\hat{S}^z = \hat{S}^z_L + \hat{S}^z_R$. Therefore, it can be 
diagonalized in the subspaces of given spin. All these subspaces are spanned together by nine natural eigenvectors of total spin 
$|\sigma,\sigma'\rangle = \hat{a}^\dagger_{L\sigma}\hat{a}^\dagger_{R\sigma'}|\mathtt{vac}\rangle$. Performing exact diagonalization of the Hamiltonian, one can find a local 
ground-state for each subspace, corresponding to a given spin. In a consequence, the true ground-state of the system is the one with the lowest energy. It is quite obvious that the states $|1,1\rangle$ and $|-1,-1\rangle$ with the largest absolute total spin ($S^z=\pm 2$) 
are the eigenstates of the Hamiltonian \eqref{eq:ham} for any set of dimensionless parameters. It comes directly from the fact that they are 
only states with the largest (smallest) total spin. Their eigenenergies are equal to $E_{\pm2} = \lambda + \tan\theta \mp 2h$, respectively. 
For these states we observe no entanglement between two sites. 

The subspaces with total spin $S^z=\pm 1$ are spanned by two natural eigenvectors: $|1,0\rangle$ and $|0,1\rangle$ for $S^z=1$, and 
$|-1,0\rangle$ and $|0,-1\rangle$ for $S^z=-1$. Due to the presence of additional $Z_2$ symmetry of the Hamiltonian \eqref{eq:ham}, namely the mirror 
{\it left-right} symmetry, the eigenstates of the Hamiltonian in these subspaces are also easy to find and they are represented by two  maximally entangled states (MES) -- Bell states. For $S^z=1$ these states are 
\begin{subequations}
\begin{align}
|\!|\mathtt{1;\pm}\rangle\!\rangle = \frac{|0,1\rangle \pm |1,0\rangle}{\sqrt{2}}
\end{align}
with corresponding eigenenergies $E_{1;\pm}=\pm(\lambda+\tan\theta)-h$.
Similarly, for $S^z=-1$ these states are 
\begin{align}
|\!|\mathtt{-1;\pm}\rangle\!\rangle = \frac{|0,-1\rangle \pm |-1,0\rangle}{\sqrt{2}}
\end{align}
\end{subequations}
and corresponding eigenenergies $E_{-1;\pm}=\pm(\lambda+\tan\theta)+h$. 
The most interesting situation is in the remaining subspace of total spin $S^z=0$. This subspace is spanned by three natural eigenvectors 
$|-1,1\rangle$, $|0,0\rangle$, and $|1,-1\rangle$. Due to the mirror symmetry of the Hamiltonian, mentioned above three eigenstates of the 
Hamiltonian in this subspace have following forms:
\begin{subequations}
\begin{align}
|\!|\mathtt{0;0}\rangle\!\rangle &=  \frac{|-1,1\rangle - |1,-1\rangle}{\sqrt{2}}, \\
|\!|\mathtt{0;\pm}\rangle\!\rangle &=\cos\alpha_\pm \frac{|-1,1\rangle + |1,-1\rangle}{\sqrt{2}} \pm \sin\alpha_\pm|0,0\rangle, 
\label{Ground00}
\end{align}
\end{subequations}
with corresponding eigenenergies:
\begin{subequations}
\begin{align}
E_{0;0} &= -\lambda+\tan\theta, \\
E_{0;\pm} &=(-\lambda+\tan\theta\pm\delta)/2,
\end{align}
\end{subequations}
where $\delta = [9(\lambda+\tan\theta)^2-4\lambda\tan\theta]^{1/2}$. The mixing angles $\alpha_\pm$ are related to the parameters of the 
Hamiltonian by the condition $\cos\alpha_\pm = \sqrt{|E_{0;\pm}|/\delta}$. It can be easily checked that the state 
$|\!|\mathtt{0;-}\rangle\!\rangle$ is a ground state in considered subspace for any set of parameters of the Hamiltonian. In addition, for 
particular set of parameters $\lambda=1$ and $\theta_0=-\pi/4$, it is degenerated with the state $|\!|\mathtt{0;0}\rangle\!\rangle$. As it 
will be explained in Sec. V, at this point the state $|\!|\mathtt{0;-}\rangle\!\rangle$ changes its character from three-dimensional triplet (for 
$\theta>\theta_0$) to three-dimensional singlet (for $\theta<\theta_0$) Bell state.

The whole spectrum of the Hamiltonian for two cases: vanishing biquadratic interaction ($\theta=0$) and vanishing external 
magnetic field ($h=0$), when positive linear interaction is assumed ($\lambda=1$), are plotted in Fig.~\ref{fig:fig1}. 
As we see there, the presence of biquadratic term does not lift the  degeneracy completely  (the remaining multiplicity of energy levels is marked by numbers in circles).
\section{Ground-state phase diagram}
\begin{figure}
\includegraphics{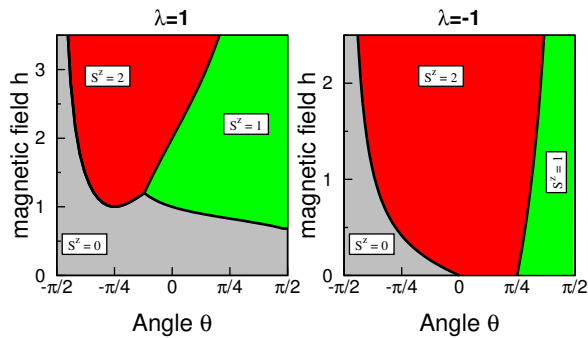}
\caption{Ground-state phase diagram of the system. For given angle $\theta$ and magnetic field $h$ ground-state of the system belongs to the 
one of three subspaces characterized with total spin $S^z$. Note that properties of the ground-state highly depend on the sign of the linear 
coupling term $\lambda$. The phase diagram for negative magnetic field $h$ has identical phases but with opposite total spin. }\label{fig2}
\end{figure}

To get more clear picture of possible scenarios we present in Fig. \ref{fig2} the ground-state phase diagram of the system. This diagram is obtained with application of exact 
diagonalization of the Hamiltonian \eqref{eq:ham}. We see that for given values of the parameters describing the Hamiltonian and magnetic filed $h>0$, the true 
ground-state of the system belongs to the one of three subspaces of total spin $S^z=0,1,2$. Obviously, for negative magnetic field $h<0$ the 
phase diagram is exactly the same but quantum phases correspond to opposite sign of the total spin. 

As it is seen from Fig.\ref{fig2}, the phase 
diagram crucially depends on the sign of the linear coupling $\lambda$. It is worth to notice that for the both cases $\lambda=\pm 1$, the 
ferrimagnetic-like quantum phase with total spin $S^z=1$ is the only phase which is confined by some minimal values of the biquadratic 
interaction. However, one should keep in mind that the ground state is always in the ferromagnetic phase (with the largest total spin $S^z=2$) for large enough values of magnetic field, whereas the 
most interesting antiferromagnetic-like phase with total spin $S^z=0$, can be reached only for sufficiently small values of magnetic field. When additionally, $\lambda$ 
is positive, this phase ($S_z=0$) can be reached for any value of biquadratic interactions. However, when $\lambda$ is negative one needs also negative 
biquadratic interaction to reach antiferromagnetic-like phase. 

Due to the fact that in the Hamiltonian \eqref{eq:ham} there is no term corresponding to the coupling states with different $S^z$, the phases are separated by 
sharp borders. All quantum phase transitions appearing in the system are of the first order -- they are characterized by discontinuities in the first derivative of the ground-
state energy. In other words, at the transition points the ground-state of the system ,,jumps'' from one subspace of total spin to the other 
one. For example, when $\lambda=1$ and $\theta=0$, we see that system undergoes two transitions for $h=1$ and for $h=2$. These transitions 
are directly related to the crossings of the energy levels shown in Fig.\ref{fig:fig1}b, and at those points magnetization of the ground-state changes its value suddenly. 

\section{Quantum correlations in the ground state}
In this section we will discuss an entanglement properties of the ground state more deeply. To study them we shall use two quantities to 
determine the entanglement present in the system. One of them is the following correlation parameter, introduced in \cite{GRE94}:
\begin{align}
{\cal K}(\hat\rho_L) = \left(\sum_i \eta_i^2 \right)^{-1},
\end{align}
where $\eta_i$ are the eigenvalues of the left-site reduced density matrix $\hat\rho_L=\mathrm{Tr}_R|\mathtt{G}\rangle\langle\mathtt{G}|$ 
obtained from the full, two-site ground state density matrix $|\mathtt{G}\rangle\langle\mathtt{G}|$ by tracing out degrees of freedom of the right-well 
(obviously this parameter can be derived for the right-site without changing the results). The parameter ${\cal K}$ gives an effective number of 
single-particle orbitals occupied in the given many-body state. In particular, when the one-site density matrix has $n$ equal eigenvalues, 
then ${\cal K} = n$. Additionally, when we are dealing with product states ${\cal K}=1$. This parameter (${\cal K}$) can be very useful is studies of bipartite systems \cite{PhysRevA.82.053631}. 

Other measure of the system entanglement, that we shall apply in this paper is von Neumann entropy. This measure is commonly used in 
numerous papers, and can be defined as
\begin{align} \label{Entropy}
{\cal S}(\hat\rho_L) = \mathrm{Tr}(\hat\rho_L \log_2\hat\rho_L) = \sum_i \eta_i\log_2\eta_i.
\end{align}
This entropy is even more interesting than the number of relevant orbitals ${\cal K}$ since it is directly related to the properties of the 
system in the thermodynamic limit, as well as in quantum information context. Entropy defined in \eqref{Entropy} for bipartite system ranges 
from $0$ for completely disentangled (product) states to $\log_2 D$ for MES defined in $D$-dimensional Hilbert space. On the other hand, we 
can discuss our system in terms of quantum information theory. For such a case, we treat the system as that of two $D$-dimensional 
\textit{qudits} \cite{RMN01}. Therefore, besides the total spin previously used, the phases can be explicitly distinguished with respect to 
the quantum state of the system and consequently, to the dimension of the Hilbert subspaces in which these states are defined. Therefore, 
the system found in a given phase can be represented by the corresponding to it quantum information theory objects. For instance, the bipartite 
system in the antiferromagnetic-like phase is described by the state defined in three-dimensional Hilbert subspace and hence, can be treated 
as \textit{qutrit-qutrit} system. Similarly, the system in ferrimagnetic-like phase can be considered as a \textit{qubit-qubit} one.

The quantities ${\cal K}$ and ${\cal S}$ characterize global properties of shared entanglement and they can be insensitive to the internal structure of the state. For example, there are various MES which are characterized with the same ${\cal K}$ and ${\cal S}$. Since we are interested in two, particular generalized Bell states \cite{sychnjp2009}:
\begin{subequations}
\begin{align}
\label{eq:3Dbellstate}
|\mathtt{3,B_S}\rangle&=\frac{1}{\sqrt{3}}\Big( |1,-1\rangle + |-1,1\rangle -|0,0\rangle\Big), \\
|\mathtt{3,B_T}\rangle&=\frac{1}{\sqrt{3}}\Big( |1,-1\rangle + |-1,1\rangle +|0,0\rangle\Big),
\end{align}
\end{subequations}
to overcome this problem we shall measure the relative distance between the ground state of the system $|\mathtt{G}\rangle$ and these Bell states.
Such relative distances are represented by the fidelities  ${\cal F}_S = |\langle \mathtt{3,B_S}|\mathtt{G}\rangle|^2$ and  ${\cal F}_T = |\langle \mathtt{3,B_T}|\mathtt{G}\rangle|^2$, respectively.

Note that, by definition, these states are not orthogonal. In consequence, even if the fidelity calculated in respect to one of them is equal to one, the remaining one is not zero. (see Fig. \ref{fig:fig6}c). Thus, only simultaneous inspection of all three parameters ($\mathcal{S}$, $\mathcal{K}$ and $\mathcal{F}_{T,S}$) can give insight into discussed physical situation. 

\subsection{Positive linear coupling ($\lambda=+1$)}
\begin{figure}
\centering
\psfrag{LABEL-FS}{\color{blue}${\cal F}_S$}
\psfrag{LABEL-FT}{\color{red}${\cal F}_T$}
\psfrag{LABEL-K}{\color{blue}${\cal K}$}
\psfrag{LABEL-Kaxis}{${\cal K}$}
\psfrag{LABEL-S}{\color{red}${\cal S}$}
\psfrag{LABEL-EntropyS}{${\cal S}$}
\psfrag{LABEL-PP}{\color{blue}$p_\pm$}
\psfrag{LABEL-P0}{\color{red}$p_0$}
\includegraphics{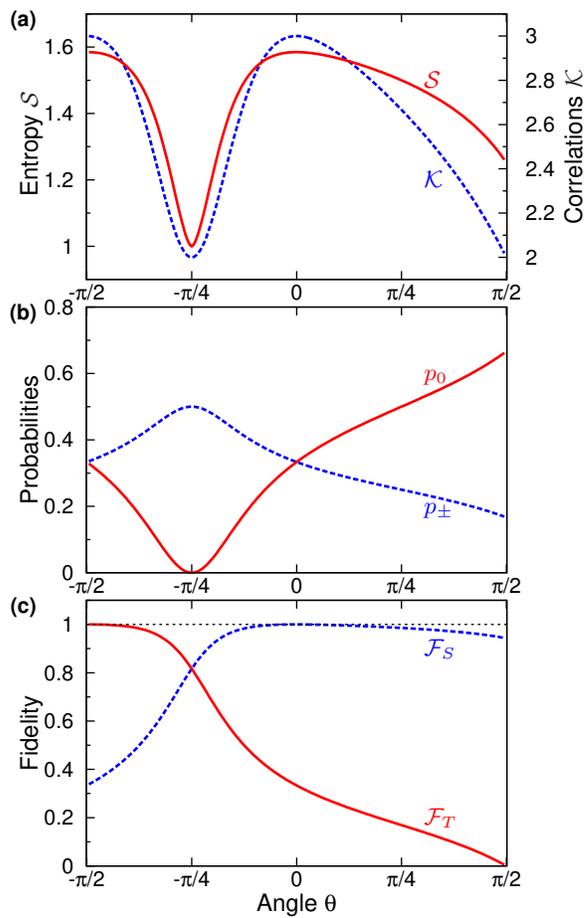}
\caption{Ground-state properties for positive linear coupling $\lambda$ in the phase with total spin $S^z=0$ as a function of the mixing 
angle $\theta$. (a) Number of relevant orbitals ${\cal K}$ (blue dashed line) and the von Neumann entropy ${\cal S}$ (red solid line), (b) probabilities $p_0$ and $p_\pm$ of finding the ground-state in appropriate two-site product states (as explained in the 
main text), and (c) fidelities ${\cal F}_S$ (${\cal F}_T$) between the ground-state of the system and singlet (triplet) {\it qutrit-qutrit} Bell state. }
\label{fig:fig6}
\end{figure}

We shall discuss various cases interesting from the point of view of the entanglement generation processes. One of the possible situations 
(the most trivial one) is that  of the  total spin $S^z=\pm 2$. Simply, being in these phases the system does not exhibit any nonlocal 
correlations. for this case ${\cal K}$ and ${\cal S}$ are equal to $1$ and $0$ respectively, and therefore, the ground-state is a product state. 

Whenever the ground-state of the system is in the ferrimagnetic-like phase (total spin $S^z=\pm 1$) the system remains in singlet Bell 
state for {\it qubit-qubit} system, $|\!|\mathtt{1,-}\rangle\!\rangle$, with ${\cal K}=2$ and entropy ${\cal S}=1$. It is worth mentioning 
that in this phase the state $|\!|\mathtt{1,-}\rangle\!\rangle$ is exact ground-state for any values of the parameters appearing in the 
Hamiltonian. Therefore, the entanglement of the ground-state in this phase is robust to any changes of these parameters. 

The most interesting situation is in the antiferromagnetic-like phase with $S^z=0$. For this case the ground-state of the system, and in 
consequence the degree of system entanglement, crucially depend on the mixing angle $\theta$. In Fig.\ref{fig:fig6}a entropy ${\cal S}$ and 
the number of relevant orbitals ${\cal K}$ for this phase are plotted. They can be calculated analytically directly from the 
equation~\eqref{Ground00} 
\begin{subequations}\label{FormulasKS}
\begin{align} 
{\cal S} =& \frac{E_{\mathtt{0;-}}}{\delta} \log_2 \left(\frac{|E_{\mathtt{0;-}}|}{2\delta} \right) + \nonumber \\
&~\frac{2\left(\lambda + \tan\theta\right)^2}{\delta~E_{\mathtt{0;-}}} 
\log_2 \left(\frac{2 \left(\lambda + \tan\theta\right)^2}{\delta~|E_{\mathtt{0;-}}|}\right), \\
{\cal K} =& \frac{2 \delta^2 E_{\mathtt{0;-}}^2}{8\left(\lambda + \tan\theta\right)^4 + E_{\mathtt{0;-}}^4}
\end{align}
\end{subequations}
The both quantities varies with the biquadratic interaction strength and hence, different {\it qutrit-qutrit} MES can be generated (Fig.\ref{fig:fig6}a). In Figs.\ref{fig:fig6}b and \ref{fig:fig6}c other quantities which allow to fully characterize properties of the ground-state 
$|\mathtt{G}\rangle$ are shown. In particular, in Fig.\ref{fig:fig6}b we plot probabilities of finding the ground-state in two site product 
states, i.e. $p_\pm= |\langle \mp 1,\pm 1|\mathtt{G}\rangle|^2$ (blue dashed line) and $p_0=|\langle 0,0|\mathtt{G}\rangle|^2$ (red solid 
line). Finally, in Fig.\ref{fig:fig6}c we present the fidelities ${\cal F}_S$ and ${\cal F}_T$.  

As it is seen from Fig.\ref{fig:fig6}a, deeply in the repulsion biquadratic interactions regime ($\theta \rightarrow -\pi/2$), the entropy 
${\cal S}=\log_23$, whereas the number of relevant orbitals ${\cal K}=3$. For such a case the ground state of the system is the maximally 
entangled triplet state
\begin{equation}
|\mathtt{G}\rangle \stackrel{^{\theta\rightarrow-\pi/2}}{\longrightarrow}|\mathtt{3,B_T}\rangle.
\end{equation}
When biquadratic interaction grows, the both: ${\cal S}$ and ${\cal K}$ start to decrease rapidly. Thus, for $\theta_0=-\pi/4$ the entropy 
${\cal S}=\log_22$ and  ${\cal K}=2$, what is related to vanishing of the probability $p_0$. 
In consequence, at this point the ground-state, initially specified in $3\otimes 3$-dimensional Hilbert space, reduces to the state 
defined in $2\otimes 2$ subspace. This state can be written as
\begin{equation}
|\mathtt{2,B_T}\rangle = \frac{1}{\sqrt{2}}\Big( |1,-1\rangle + |-1,1\rangle \Big).
\end{equation}
In fact, it is Bell MES defined in $2\otimes 2$ Hilbert space, and we deal here with a \textit{qubit-qubit} system. 
Moreover, one should remember that  for $\theta=\theta_0$ the ground-state $|\mathtt{2,B_T}\rangle$ is degenerate with other Bell state 
$|\!|\mathtt{0,0}\rangle\!\rangle$.

When $\theta>\theta_0$ the probability amplitude $\langle 0,0|\mathtt{G}\rangle$ changes its sign  and in consequence, the ground-state 
changes its nature, switching form triplet-like to singlet-like state (they are not perfect triplet and singlet states, as some amount of 
the probability corresponding to other states is present in the system). Such switching can be realized in practical realizations for 
instance, by adiabatic changes of the parameters of the Hamiltonian. Moreover, from Fig.\ref{fig:fig6}c we see that the fidelity 
${\cal F}_S$ increases and becomes larger than ${\cal F}_T$ for $\theta>\theta_0$ and then, reaches its maximal value (${\cal F}_S = 1$) 
for vanishing biquadratic interaction ($\theta=0$). At this point the singlet Bell state is generated
\begin{equation}
|\mathtt{G}\rangle \stackrel{^{\theta=0}}{=\!\!=}|\mathtt{3,B_S}\rangle ,
\end{equation}
what is manifested by ${\cal S}=\log_23$ and ${\cal K}=3$.

For attractive interactions ($\theta>0$) the ground-state remains almost exactly in the singlet  state $|\mathtt{3,B_S}\rangle$, 
\textit{i.e.} the fidelity ${\cal F}_S$ decreases but not more than $\sim 10\%$ from the unity. Moreover, we see that for this case the 
entropy decreases, the same as the number of relevant orbitals. In the limit $\theta \rightarrow \pi/2$ we have ${\cal K}=2$. This fact 
shows that only two relevant orbitals are involved, and might suggest that we are dealing with the same situation as that for 
$\theta =\theta_0$. However, for this situation the entropy $\log_23>{\cal S}>\log_22$. In consequence, although the state of our system is defined in $3\otimes 3$ 
Hilbert space, it is not MES. The fact that ${\cal K}=2$ is caused by the distribution of the probabilities $p_0$ and $p_\pm$. As we can see from Fig.\ref{fig:fig6}b, for such situation the probability $p_0$ plays a dominant role and is four times higher then $p_\pm$. Therefore, the sum of the both probability amplitudes, corresponding to $p_\pm$, is equal to the amplitude corresponding to $p_0$. In consequence, the  effective number of relevant orbitals is equal to 2.
 
\subsection{Negative linear coupling ($\lambda=-1$)}
\begin{figure}
\centering
\psfrag{LABEL-FS}{\color{blue}${\cal F}_T$}
\psfrag{LABEL-FT}{\color{red}${\cal F}_S$}
\psfrag{LABEL-K}{\color{blue}${\cal K}$}
\psfrag{LABEL-Kaxis}{${\cal K}$}
\psfrag{LABEL-S}{\color{red}${\cal S}$}
\psfrag{LABEL-EntropyS}{${\cal S}$}
\psfrag{LABEL-PP}{\color{blue}$p_\pm$}
\psfrag{LABEL-P0}{\color{red}$p_0$}
\includegraphics{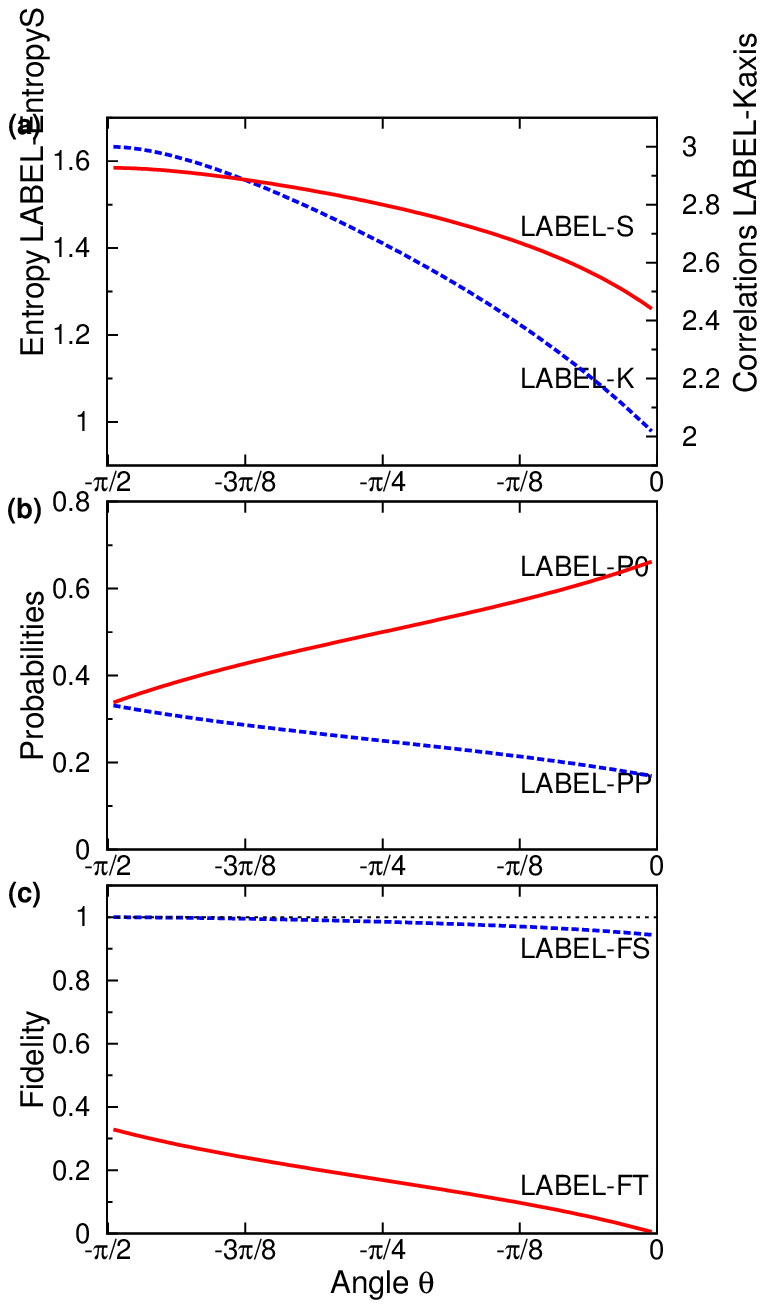}
\caption{Ground-state properties for negative linear coupling $\lambda$ in the phase with total spin $S^z=0$ as a function of the mixing 
angle $\theta$. (a) Number of relevant orbitals ${\cal K}$ (blue dashed line) and the von Neumann entropy ${\cal S}$ (red solid line), 
(b) Probabilities $p_0$ and $p_\pm$ of finding the ground-state in appropriate two-site product states (as explained in the main text), (c) 
Fidelities ${\cal F}_S$ and ${\cal F}_T$ between the ground-state of the system and singlet and triplet {\it qutrit-qutrit} Bell state 
respectively. In contrast to the case $\lambda=1$, the ground-state remains almost perfectly in the triplet {\it qutrit-qutrit} Bell state 
$|\mathtt{3,B_T}\rangle$ in a whole phase.}
\label{fig:fig7}
\end{figure}

Similarly, correlations in the ground-state can be discussed for negative linear coupling $\lambda=-1$. As previously, for the phases characterized by 
total spin $S^z=2$ and $S^z=1$, the ground-state is product $|1,1\rangle$ and singlet {\it qubit-qubit} Bell state 
$|\!|\mathtt{1,-}\rangle\!\rangle$, respectively.

Situation changes for the case when total spin $S^z=0$, what can be observed only for negative values of the angle $\theta$. For such a situation all 
properties of the ground state can also be derived analytically. In particular, the formulas for the entropy $\cal S$ and the number of 
relevant orbitals ${\cal K}$ have the same form  in eq. \eqref{FormulasKS}. They are plotted in Fig.\ref{fig:fig7}a. As it is seen from 
Fig.\ref{fig:fig7}b, in contrast to the case $\lambda=1$, for negative linear coupling the ground-state of the system remains almost perfectly in the triplet {\it qutrit-qutrit} Bell state $|\mathtt{3,B_T}\rangle$. In the limit of infinite repulsive biquadratic 
interactions, the fidelity ${\cal F}_T$ is equal to 1. Moreover, all three probabilities $p_\pm$ and $p_0$ are equal to $1/3$. All these means that 
for the cases when $\lambda=-1$, the degree of bipartite entanglement for the ground state is highly insensitive on the values of the parameters 
describing the superlattice. One should remember that such robustness appears  when the system is in the phase with vanishing total spin 
$S^z=0$, and was not present for $\lambda=1$ (see the cases discussed in previous section).  

\section{Conclusions}

We have discussed the model of two spin-1 bosons confined in the double-well potential of the optical superlattice and influenced by an 
external magnetic field. We showed that dependent on the values of the parameters appearing in the Hamiltonian, the ground-state can belong 
to different phases distinguished by total spin $S^z$, and the presence of non-linear biquadratic interaction considerably influences the 
system's properties. For experimentally accessible values of the parameters the ferro-, ferri- and antiferromagnetic-like phases have been 
identified and presented as a separated sectors in phase-diagrams. We have pointed out that the sharp boundaries between 
sectors are related to the crossings of energies in the spectrum of the Hamiltonian. Subsequently, we have uniquely determined ground-state of the system of a given magnetization for each magnetic phase. In this paper we have concentrated on the possibility of generation of 
MES defined in $2\otimes 2$ and $3\otimes 3$ Hilbert spaces, and we have identified the set of parameters for which MES of different kinds can be achieved. What is important, applying changes in the parameters describing  optical lattice, we can switch the system 
from one MES to another. For instance, we can continuously transform the state of our system from a singlet to a triplet Bell state. 
Moreover, it is possible to change the system's state from MES defined in $2\otimes 2$ Hilbert subspace to that, defined in $3\otimes 3$-dimensional space, and \textit{vice versa}. Such switching performed within a single phase can be induced by adiabatic changes of the 
parameters of the Hamiltonian. For the cases when transitions are made between two phases, some additional interaction should be involved. 
Such interaction (for instance, interaction with external bath) is necessary, as the system has to change its total spin which commutes with the Hamiltonian.

In particular, we have found that in the ferrimagnetic-like phase the ground-state is a \textit{qubit-qubit} singlet Bell state regardless 
of the type of the Heisenberg interaction. Moreover, discussing the case of antiferromagnetic-like phase, we have showed that the 
biquadratic term of Heisenberg interactions plays a crucial role in determining properties of the ground-state of the system. For 
$\lambda=-1$ the ground-state is the \textit{qutrit-qutrit} triplet Bell state while for $\lambda=1$ the continuous transition from the 
\textit{qutrit-qutrit} singlet to \textit{qutrit-qutrit} triplet can be induced by adiabatic varying of biquadratic interaction. Moreover, at 
the transition point, the ground-state becomes the \textit{qubit-qubit} triplet Bell state.

We believe that the system studied here can be a potential candidate for practical realization of a  device which could be applied as a switchable tool for generation of various MES on demand.

\begin{acknowledgments}
We would like to thank prof. A. Miranowicz for his valuable suggestions and discussions. 
The authors wish to thank the (Polish) National Science Center (grant No. DEC-2011/01/D/ST2/02019) for the support. Numerical calculations were performed in WCSS Wroc{\l}aw (Poland). 

\end{acknowledgments}


\begin{thebibliography}{10}
\newcommand{\enquote}[1]{``#1''}

\bibitem{bennetnatere2000}
C.~H. Bennett and D.~P. DiVincenzo, \enquote{Quantum information and
  computation,} Nature (London) \textbf{404}, 247 (2000).

\bibitem{boschiprl1998}
D.~Boschi, S.~Branca, F.~{De Martini}, L.~Hardy, and S.~Popescu,
  \enquote{Experimental realization of teleporting an unknown pure quantum
  state via dual classical and einstein-podolsky-rosen channels,} Phys. Rev.
  Lett. \textbf{80}, 1121 (1998).

\bibitem{bouwmeesternaure1997}
D.~Bouwmeester, J.~W. Pan, K.~Mattle, M.~Eible, H.~Weinfurter, and
  A.~Zeilinger, \enquote{Experimental quantum teleportation,} Nature
  \textbf{390}, 575 (1997).

\bibitem{M05}
A.~Miranowicz, \enquote{Optical-state truncation and teleportation of qudits by
  conditional eight-port interferometry,} J. Opt. B: Quant. Semiclass. Opt.
  \textbf{7}, 142 (2005).

\bibitem{OKB07}
S.~K. Ozdemir, K.~Bartkiewicz, Y.~X. Liu, and A.~Miranowicz,
  \enquote{Teleportation of qubit states through dissipative channels:
  Conditions for surpassing the no-cloning limit,} Phys. Rev. A \textbf{76},
  042325 (2007).

\bibitem{GK13}
S.~K. Goyal and T.~Konrad, \enquote{Teleporting photonic qudits using multimode
  quantum scissors,} Scientific Reports \textbf{3}, 3548 (2013).

\bibitem{gisinrmp2002}
N.~Gisin, G.~Ribordy, W.~Tittel, and H.~Zbinden, \enquote{Quantum
  cryptography,} Rev. Mod. Phys. \textbf{74}, 145 (2002).

\bibitem{BLC13}
K.~Bartkiewicz, K.~Lemr, A.~Cernoch, J.~Soubusta, and A.~Miranowicz,
  \enquote{Experimental eavesdropping based on optimal quantum cloning,} Phys.
  Rev. Lett. \textbf{110}, 173601 (2013).

\bibitem{levineprl2004}
G.~C. Levine, \enquote{Entanglement entropy in a boundary impurity model,}
  Phys. Rev. Lett. \textbf{93}, 266402 (2004).

\bibitem{kitaevprl2006}
A.~Kitaev and J.~Preskill, \enquote{Topological entanglement entropy,} Phys.
  Rev. Lett. \textbf{96}, 110404 (2006).

\bibitem{BMW11}
M.~Bartkowiak, A.~Miranowicz, X.~Wang, Y.~X. Liu, W.~Leo\'nski, and F.~Nori,
  \enquote{Sudden vanishing and reappearance of nonclassical effects: General
  occurrence of finite-time decays and periodic vanishings of nonclassicality
  and entanglement witnesses,} Phys. Rev. A \textbf{83}, 053814 (2011).

\bibitem{ghoshnature2003}
S.~Ghosh, T.~F. Rosenbaum, G.~Aeppli, and S.~N. Coppersmith, \enquote{Entangled
  quantum state of magnetic dipoles,} Nature \textbf{425}, 48 (2003).

\bibitem{osterlohnature2002}
A.~Osterloh, L.~Amico, G.~Falci, and R.~Fazio, \enquote{Scaling of entanglement
  close to a quantum phase transition,} Nature \textbf{416}, 608 (2002).

\bibitem{osbornepra2002}
T.~J. Osborne and M.~A. Nielsen, \enquote{Entanglement in a simple quantum
  phase transition,} Phys. Rev. A \textbf{66}, 032110 (2002).

\bibitem{amicormp2008}
L.~Amico, R.~Fazio, A.~Osterloh, and V.~Vedral., \enquote{Entanglement in
  many-body systems,} Rev. Mod. Phys. \textbf{80}, 517 (2008).

\bibitem{pengpra2005}
X.~Peng, J.~Du, and D.~Suter, \enquote{Quantum phase transition of ground-state
  entanglement in a heisenberg spin chain simulated in an nmr quantum
  computer,} Phys. Rev. A \textbf{71}, 012307 (2005).

\bibitem{zhangprl2008}
J.~Zhang, X.~Peng, N.~Rajendran, and D.~Suter, \enquote{Detection of quantum
  critical points by a probe qubit,} Phys. Rev. Lett. \textbf{100}, 100501
  (2008).

\bibitem{leweoxford2012}
M.~Lewenstein, A.~Sanpera, and V.~Ahufinger, \emph{Ultracold atoms in optical
  lattices: Simulating quantum many-body systems} (Oxford University Press,
  Oxford, UK,, 2012).

\bibitem{blattnatphys2012}
R.~Blatt and C.~F. Roos, \enquote{Quantum simulations with trapped ions,} Nat.
  Phys. \textbf{8}, 277 (2012).

\bibitem{blochnatphys2012}
I.~Bloch, J.~Dalibard, and S.~Nascimbene, \enquote{Quantum simulations with
  ultracold quantum gases,} Nat. Phys. \textbf{8}, 267 (2012).

\bibitem{losspra1998}
D.~Loss and D.~P. DiVincenzo, \enquote{Quantum computation with quantum dots,}
  Phys. Rev. A \textbf{57}, 120 (1998).

\bibitem{greinernature2002}
M.~Greiner, O.~Mandel, T.~Esslinger, T.~W. Hansch, and I.~Bloch,
  \enquote{Quantum phase transition from a superfluid to a mott insulator in a
  gas of ultracold atoms,} Nature \textbf{415}, 39 (2002).

\bibitem{blochrevmodphys2008}
I.~Bloch, J.~Dalibard, and W.~Zwerger, \enquote{Many-body physics with
  ultracold gases,} Rev. Mod. Phys. \textbf{80}, 885 (2008).

\bibitem{hungnature2011}
C.~Hung, X.~Zhang, N.~Gemelke, and C.~Chin, \enquote{Observation of scale
  invariance and universality in two-dimensional bose gases,} Nature
  \textbf{470}, 236 (2011).

\bibitem{nikolopoulosepl65_2004}
G.~M. Nikolopoulos, D.~Petrosyan, and P.~Lambropoulos, \enquote{Coherent
  electron wavepacket propagation and entanglement in array of coupled quantum
  dots,} Europhys. Lett. \textbf{65}, 297 (2004).

\bibitem{christandlprl92_2004}
M.~Christandl, N.~Datta, A.~Ekert, and A.~J. Landahl, \enquote{Perfect state
  transfer in quantum spin networks,} Phys. Rev. Lett. \textbf{92}, 187902
  (2004).

\bibitem{wupra80_2009}
L.-A. Wu, A.~Miranowicz, X.~B. Wang, Y.~X. Liu, and F.~Nori, \enquote{Perfect
  function transfer and interference effects in interacting boson lattices,}
  Phys. Rev. A \textbf{80}, 012332 (2009).

\bibitem{weitenbergnature2011}
C.~Weitenberg, M.~Endres, J.~F. Sherson, M.~Cheneau, P.~Schau\ss, T.~Fukuhara,
  I.~Bloch, and S.~Kuhr, \enquote{Single-spin addressing in an atomic mott
  insulator,} Nature \textbf{471}, 319 (2011).

\bibitem{pietraszkiewiczpra88_2013}
J.~Pietraszewicz, T.~Sowi\'nski, M.~Brewczyk, M.~Lewenstein, and M.~Gajda,
  \enquote{Spin dynamics of two bosons in an optical lattice site: A role of
  anharmonicity and anisotropy of the trapping potential,} Phys. Rev. A
  \textbf{88}, 013608 (2013).

\bibitem{Feynman}
R.~P. Feynman, \enquote{Simulating physics with computers,} Int. J. Theor.
  Phys. \textbf{21}, 467 (1982).

\bibitem{Hauke}
P.~Hauke, F.~M. Cucchietti, L.~Tagliacozzo, I.~Deutsch, and M.~Lewenstein,
  \enquote{Can one trust quantum simulators?} Rep. Prog. Phys. \textbf{75},
  082401 (2012).

\bibitem{jakschprl81_1998}
D.~Jaksch, C.~Bruder, J.~I. Cirac, C.~W. Gardiner, and P.~Zoller, \enquote{Cold
  bosonic atoms in optical lattices,} Phys. Rev. Lett. \textbf{81}, 3108
  (1998).

\bibitem{PhysRevLett.95.030405}
O.~E. Alon, A.~I. Streltsov, and L.~S. Cederbaum, \enquote{Zoo of quantum
  phases and excitations of cold bosonic atoms in optical lattices,} Phys. Rev.
  Lett. \textbf{95}, 030405 (2005).

\bibitem{PhysRevLett.95.033003}
V.~W. Scarola and S.~{Das Sarma}, \enquote{Quantum phases of the extended
  bose-hubbard hamiltonian: Possibility of a supersolid state of cold atoms in
  optical lattices,} Phys. Rev. Lett. \textbf{95}, 033003 (2005).

\bibitem{PhysRevA.85.033638}
F.~Pinheiro, J.-P. Martikainen, and J.~Larson, \enquote{Confined p-band
  bose-einstein condensates,} Phys. Rev. A \textbf{85}, 033638 (2012).

\bibitem{PhysRevLett.111.215302}
T.~Sowi\'nski, M.~\L{}\c{a}cki, O.~Dutta, J.~Pietraszewicz, P.~Sierant,
  M.~Gajda, J.~Zakrzewski, and M.~Lewenstein, \enquote{Tunneling-induced
  restoration of the degeneracy and the time-reversal symmetry breaking in
  optical lattices,} Phys. Rev. Lett. \textbf{111}, 215302 (2013).

\bibitem{PhysRevLett.108.165301}
T.~Sowi\'nski, \enquote{Creation on demand of higher orbital states in a
  vibrating optical lattice,} Phys. Rev. Lett. \textbf{108}, 165301 (2012).

\bibitem{auerbach1994}
A.~Auerbach, \emph{Interacting Electrons and Quantum Magnetism}
  (Springer-Verlag, Berlin, German, 1994).

\bibitem{simonsnature2011}
J.~Simon, W.~S. Bakr, R.~Ma, M.~E. Tai, P.~M. Preiss, and M.~Greiner,
  \enquote{Quantum simulation of antiferromagnetic spin chains in an optical
  lattice,} Nature \textbf{472}, 307 (2011).

\bibitem{andersonpr1959}
P.~W. Anderson, \enquote{New approach to the theory of superexchange
  interactions,} Phys. Rev. \textbf{115}, 2 (1959).

\bibitem{imambekovpra68_2003}
A.~Imambekov, M.~Lukin, and E.~Demler, \enquote{Spin-exchange interactions of
  spin-one bosons in optical lattices: Singlet, nematic, and dimerized phases,}
  Phys. Rev. A \textbf{68}, 063602 (2003).

\bibitem{yipplr2003}
S.-K. Yip, \enquote{Dimer state of spin-1 bosons in an optical lattice,} Phys.
  Rev. Lett. \textbf{90}, 250402 (2003).

\bibitem{wumplb20_2006}
C.~Wu, \enquote{Hidden symmetry and quantum phases in spin-3/2 cold atomic
  systems,} Mod. Phys. Lett. B \textbf{20}, 1707 (2006).

\bibitem{eckertnjp9_2007}
K.~Eckert, L.~Zawitkowski, M.~J. Leskinen, A.~Sanpera, and M.~Lewenstein,
  \enquote{Ultracold atomic bose and fermi spinor gases in optical lattices,}
  New J. Phys. \textbf{9}, 133 (2007).

\bibitem{hermeleprl103_2009}
M.~Hermele, V.~Gurarie, and A.~M. Rey, \enquote{Mott insulators of ultracold
  fermionic alkaline earth atoms: Underconstrained magnetism and chiral spin
  liquid,} Phys. Rev. Lett. \textbf{103}, 135301 (2009).

\bibitem{garciaprl2004}
J.~J. Garcia-Ripoll, M.~A. Martin-Delgado, and J.~I. Cirac,
  \enquote{Implementation of spin hamiltonians in optical lattices,} Phys. Rev.
  Lett. \textbf{93}, 250405 (2004).

\bibitem{DL94}
A.~Drzewi\'nski and J.~M.~J. {van Leeuwen}, \enquote{Renormalization of the
  ising model in a transverse field,} Phys. Rev. B \textbf{49}, 403 (1994).

\bibitem{DD95}
A.~Drzewi\'nski and R.~Dekeyser, \enquote{Renormalization of the anisotropic
  linear xy model,} Phys. Rev. B \textbf{51}, 15218 (1995).

\bibitem{zhoupra2003}
L.~Zhou, H.~S. Song, Y.~Q. Guo, and C.~Li, \enquote{Enhanced thermal
  entanglement in an anisotropic heisenberg xyz chain,} Phys Rev. A
  \textbf{68}, 024301 (2003).

\bibitem{hiranojpsj2007}
T.~Hirano and Y.~Hatsugai, \enquote{Entanglement entropy of one-dimensional
  gapped spin chains,} J. Phys. Soc. Jpn. \textbf{76}, 074603 (2007).

\bibitem{pendpra2010}
X.~Peng, J.~Zhang, J.~Du, and D.~Suter, \enquote{Ground-state entanglement in a
  system with many-body interactions,} Phys. Rev. A \textbf{81}, 042327 (2010).

\bibitem{guoepjd2010}
J.~L. Guo and H.~S. Song, \enquote{Entanglement and teleportation through a
  two-qubit heisenberg xxz model with the dzyaloshinskii-moriya interaction,}
  Eur. Phys. J. D \textbf{56}, 265 (2010).

\bibitem{rodrigezprl2011}
K.~Rodriguez, A.~Arg\"uelles, A.~K. Kolezhuk, L.~Santos, and T.Vekua,
  \enquote{Field-induced phase transitions of repulsive spin-1 bosons in
  optical lattices,} Phys. Rev. Lett. \textbf{106}, 105302 (2011).

\bibitem{chiaraprb2011}
G.~De~Chiara, M.~Lewenstein, and A.~Sanpera, \enquote{Bilinear-biquadratic
  spin-1 chain undergoing quadratic zeeman effect,} Phys. Rev. B \textbf{84},
  054451 (2011).

\bibitem{yippra85_2012}
P.~Chen, Z.-L. Xue, I.~P. McCulloch, M.-C. Chung, and S.-K. Yip,
  \enquote{Dimerized and trimerized phases for spin-2 bosons in a
  one-dimensional optical lattice,} Phys. Rev. A \textbf{85}, 011601(R) (2012).

\bibitem{millet1999}
P.~Millet, F.~Mila, F.~C. Zhang, M.~Mambrini, A.~B. {Van Oosten}, V.~A.
  Pashchenko, A.~Sulpice, and A.~Stepanov, \enquote{Biquadratic interactions
  and spin-peierls transition in the spin-1 chain {LiVGe}$_2${O}$_6$,} Phys.
  Rev. Lett. \textbf{83}, 4176 (1999).

\bibitem{louprl2000}
J.~Lou, T.~Xiang, and Z.~Su, \enquote{Thermodynamics of the
  bilinear-biquadratic spin-one heisenberg chain,} Phys. Rev. Lett.
  \textbf{85}, 2380 (2000).

\bibitem{basterdisprb2007}
R.~Bastardis, N.~Guih\`ery, and C.~de~Graaf, \enquote{Microscopic origin of
  isotropic non-heisenberg behavior in $s=1$ magnetic systems,} Phys. Rev. B
  \textbf{76}, 132412 (2007).

\bibitem{benciniica2008}
A.~Bencini and F.~Totti, \enquote{On the importance of the biquadratic terms in
  exchange coupled systems: A post-hf investigation,} Inorganica Chimica Acta
  \textbf{361}, 4153 (2008).

\bibitem{semenkainchem2010}
V.~V. Semenaka, O.~V. Nesterova, V.~N. Kokozay, V.~V. Dyakonenko, R.~I.
  Zubatyuk, O.~Shishkin, R.~Bo\v{c}a, J.~Jezierska, and A.~Ozarowski,
  \enquote{{Cr}$^{III}-${Cr}$^{III}$ interactions in two alkoxo-bridged
  heterometallic {Zn}$_2${Cr}$_2$ complexes self-assembled from zinc oxide,
  reineckes salt, and diethanolamine,} Inorg. Chem. \textbf{49}, 5460
  (2010).

\bibitem{wagnerpra84_2011}
A.~Wagner, C.~Bruder, and E.~Demler, \enquote{Spin-1 atoms in optical
  superlattices: Single-atom tunneling and entanglement,} Phys. Rev. A
  \textbf{84}, 063636 (2011).

\bibitem{papoularpra81_2010}
D.~J. Papoular, G.~V. Shlyapnikov, and J.~Dalibard, \enquote{Microwave-induced
  fano-feshbach resonances,} Phys. Rev. A \textbf{81}, 041603(R) (2010).

\bibitem{LMG10}
Y.~X. Liu, A.~Miranowicz, Y.~B. Gao, J.~J.~Bajer, C.~P. Sun, and F.~Nori,
  \enquote{Qubit-induced phonon blockade as a signature of quantum behavior in
  nanomechanical resonators,} Phys. Rev. A \textbf{82}, 032101 (2010).

\bibitem{GSK12}
T.~V. Gevorgyan, A.~R. Shahinyan, and G.~Y. Kryuchkyan, \enquote{Generation of
  fock states and qubits in periodically pulsed nonlinear oscillators,} Phys.
  Rev. A \textbf{85}, 053802 (2012).

\bibitem{MPL13}
A.~Miranowicz, M.~Paprzycka, Y.~X. Liu, J.~Bajer, and F.~Nori,
  \enquote{{Two-photon and three-photon blockades in driven nonlinear
  systems},} {Phys. Rev. A} \textbf{87}, 023809 (2013).

\bibitem{LT94}
W.~Leo\'nski and R.~Tana\'s, \enquote{Possibility of producing the one-photon
  state in a kicked cavity with a nonlinear kerr medium,} Phys. Rev. A
  \textbf{49}, R20 (1994).

\bibitem{ML04}
A.~Miranowicz and W.~Leo\'nski, \enquote{Dissipation in systems of linear and
  nonlinear quantum scissors,} Journal of Optics B: Quant. Semiclass. Opt.
  \textbf{6}, S43 (2004).

\bibitem{KL10}
A.~Kowalewska-Kud{\l}aszyk and W.~Leo\'nski, \enquote{Squeezed vacuum reservoir
  effect for entanglement decay in the nonlinear quantum scissor system,} J.
  Phys. B: At. Mol. Opt. Phys. \textbf{43}, 205503 (2010).

\bibitem{LK11}
W.~Leo\'nski and A.~Kowalewska-Kud{\l}aszyk, \enquote{Quantum scissors 
  finite-dimensional states engineering,} Progress in Optics \textbf{56}, 131
  (2011).

\bibitem{M90}
G.~J. Milburn, \enquote{Coherence and chaos in a quantum optical system,} Phys.
  Rev. A \textbf{41}, 6567 (1990).

\bibitem{MH91}
G.~J. Milburn and C.~A. Holmes, \enquote{Quantum coherence and classical chaos
  in a pulsed parametric oscillator with a kerr nonlinearity,} Phys. Rev. A
  \textbf{44}, 4704 (1991).

\bibitem{L96}
W.~Leo\'nski, \enquote{Quantum and classical dynamics for a pulsed nonlinear
  oscillator,} Physica A: Statistical Mechanics and its Applications
  \textbf{233}, 365 (1996).

\bibitem{KKL09}
A.~Kowalewska-Kud{\l}aszyk, J.~K. Kalaga, and W.~Leo\'nski, \enquote{Long-time
  fidelity and chaos for a kicked nonlinear oscillator system,} Phys. Lett. A
  \textbf{373}, 1334 (2009).

\bibitem{GSC13}
T.~V. Gevorgyan, A.~R. Shahinyan, L.~Y. Chew, and G.~Y. Kryuchkyan,
  \enquote{Bistability and chaos at low levels of quanta,} Phys. Rev. E
  \textbf{88}, 022910 (2013).

\bibitem{GRE94}
R.~Grobe, K.~Rz\c{a}\.zewski, and J.~H. Eberly, \enquote{Measure of
  electron-electron correlations in atomic physics,} J. Phys. B \textbf{27},
  L503 (1994).

\bibitem{PhysRevA.82.053631}
T.~Sowi\'nski, M.~Brewczyk, M.~Gajda, and K.~Rz\c{a}\.zewski, \enquote{Dynamics
  and decoherence of two cold bosons in a one-dimensional harmonic trap,} Phys.
  Rev. A \textbf{82}, 053631 (2010).

\bibitem{RMN01}
P.~Rungta, W.~J. Munro, K.~Nemoto, P.~Deuar, G.~J. Milburn, and C.~M. Caves,
  \enquote{{Qudit entanglement},} in \enquote{{DIRECTIONS IN QUANTUM OPTICS},}
  , vol. {561} of \emph{{LECTURE NOTES IN PHYSICS}}, {H.~J.~Carmichael and
  R.~J.~Glauber and M.~O.~Scully}, ed. ({2001}), vol. {561} of \emph{{LECTURE
  NOTES IN PHYSICS}}, pp. {149--164}. {TAMU-ONR Workshop on Quantum Optics,
  JACKSON, WY, JUL 26-30, 1999}.

\bibitem{sychnjp2009}
D.~Sych and G.~Leuchs, \enquote{A complete basis of generalized bell states,}
  New J. Phys. \textbf{11}, 013006 (2009).

\end{thebibliography}
\end{document}